\documentclass[aps, prb, floatfix, twocolumn, reprint]{revtex4-1}

\usepackage{graphicx}
\usepackage{amsmath}
\usepackage{color}
\usepackage{hyperref}

\hypersetup{
	colorlinks = true, 
	urlcolor   = blue, 
	linkcolor  = blue, 
	citecolor  = blue,   
}

\begin{document}
	\title{Driving Skyrmions in a Composite Bilayer}
	\author{Zidong Wang}
	\email{Zidong.Wang@auckland.ac.nz}
	\author{Malcolm J. Grimson}
	\email{m.grimson@auckland.ac.nz}
	\affiliation{Department of Physics, University of Auckland, Private Bag 92019, Auckland, New Zealand}
	\date{\today}

\begin{abstract}
	Magnetic Skyrmions and multiferroics are the most interesting objects in nanostructure science that have great potential in future spin-electronic technology. The study of multiferroic Skyrmions has attracted much interest in recent years. This Article reports magnetic Bloch Skyrmions induced by an electric driving field in a composite bilayer (chiral-magnetic/ferroelectric bilayer) lattice. By using the spin dynamics method, we use a classical magnetic spin model and an electric pseudospin model, which are coupled by a strong magnetoelectric coupling in the dynamical simulations. Interestingly, we observe some skyrmion-like objects in the electric component either during the switching process or by applying a magnetic field, which is due to the connection between the electric and the magnetic structures.\\
	\\
	\noindent \url{http://dx.doi.org/10.1103/PhysRevB.94.014311}
\end{abstract}

\maketitle

\section{Introduction}

The Skyrmion is a topological particle-like object, named after Skyrme, who described this in quantum field theory in the 1960's \cite{Nucl.Phys.31.556}. Decades later, with the emergence of spintronics, Bogdanov and R{\"o}{\ss}ler \cite{N.442.797}, successfully predicted that Skyrmions can be induced by an inhomogeneous magnetic chiral interaction (i.e., the Dzyaloshinskii-Moriya interaction). The Dzyaloshinskii-Moriya interaction characterizes the asymmetric exchange interaction between the magnetic spin and its neighbors. It is a key ingredient to break chiral symmetry in magnetic nanostructures \cite{PRL.87.037203}. Consequently, the Skyrmion has an asymmetric spiral-like topological spin texture and it offers numerous advantages for potential spin-electronic technology. Magnetic Skyrmions have been found in metallic \textit{B20} materials, such as ${\text{MnSi}}$ \cite{S.323.915}, ${\text{FeGe}}$ \cite{NM.10.106}, $ {\text{Fe}}_{1-x}{\text{Co}}_x{\text{Si}} $ \cite{N.465.901}, $ {\text{MnGe}} $ and $ {\text{Mn}}_{1-x}{\text{Fe}}_x{\text{Ge}} $ \cite{NN.8.723}, which belong to the chiral magnets. One application of magnetic Skyrmions is to provide the low-energy cost of writing, reading, and erasing non-volatile memory \cite{NC.7.10293}. 

Multiferroics offer the possibility of electric-induced magnetization and magnetic-induced (electric) polarization due to the magnetoelectric coupling between the magnetic dipoles and the electric dipoles \cite{N.442.759}. Many experiments \cite{N.430.541, PRB.88.121409, APL.106.262902} have proved the existence of this multiferroism. Furthermore, two types of multiferroic materials have been discovered, the single phase and the composite phase \cite{N.442.759}. Single-phase multiferroics have received intense investigation in recent years. Seki \textit{et al.} have observed magnetic Skyrmions controlled by an external electric field in the ${\text{Cu}}_{2}{\text{OSeO}}_{3}$ crystal lattice \cite{S.336.198}. Generally, it is possible to write magnetoelectric Skyrmions induced by electric polarization in an insulating chiral multiferroic \cite{PRB.87.100402, SR.5.8318, APL.107.082409, JPC.27.503001}. However, as a single-phase multiferroic, ${\text{Cu}}_{2}{\text{OSeO}}_{3}$ has a weak magnetic response, and its multiferroism only works at a low transition temperature, which are adverse for applications \cite{PRB.78.094416}. The other category is composite multiferroics, which is an artificially synthesized heterostructure of two materials, one a magnet and the other a ferroelectric \cite{N.442.759}. Composite multiferroics have remarkable magnetoelectric coupling due to their indirect strain-stress effect \cite{PRB.50.6082}.

In this respect, we propose a model of a composite bilayer with a heterostructure of a chiral-magnetic (CM) layer and a ferroelectric (FE) layer. Magnetoelectric coupling is used to generate multiferroic Skyrmions in this work. So far, the classical spin model is generally used to describe the behavior of  magnetization in magnets \cite{PRB.55.12556}. The spatial distribution of the magnetization $ \textbf{M} = (M_x , M_y , M_z) $ is given by a reduced magnetization $ \textbf{S} = \textbf{M} / M_s $, where $ M_s $ is the saturated magnetization. This is called the magnetic spin. The dynamics in the composite bilayer involves a microscopic study of electric properties. Generally, the behaviors of the electric polarization are described by the Landau-Devonshire phenomenological theory \cite{SSC.49.823}. But this case, the phenomenological theory has a different length scale to the micromagnetics. Hence we use a pseudospin model to investigate the energy in the FE structure \cite{EPJAP.70.30303}. This model was introduced by de Gennes \cite{SSC.1.132} and Elliott \textit{et al.} \cite{FE.7.23}. We have extended it to deal with this system. The electric pseudospin $ \textbf{P} = (P_x , P_y , P_z)  $ is a polarization vector, but only the \textit{z}-component of the pseudospins contributes to the energy in the model Hamiltonian.

In this Article, we explore magnetic Bloch Skyrmions and find electric \textquotedblleft footprint Skyrmions\textquotedblright~in a composite bilayer lattice by using the Landau-Lifshitz equations numerically. In \textbf{Section~\ref{Model}}, the model of a CM/FE stacked structure is introduced. The spin dynamics method is described in \textbf{Section~\ref{Method}}. Electric-field-induced magnetic Skyrmions and electric \textquotedblleft footprint Skyrmions\textquotedblright~are detailed in \textbf{Section~\ref{E-Skyrmions}}. \textbf{Section~\ref{M-Skyrmions}} demonstrates magnetic-field-induced electric \textquotedblleft footprint Skyrmions\textquotedblright. The paper concludes with a discussion in \textbf{Section~\ref{Conclusion}}.

\section{Model}
\label{Model}

\begin{figure}
	\includegraphics[width=250px, trim=0 400 0 300, clip]{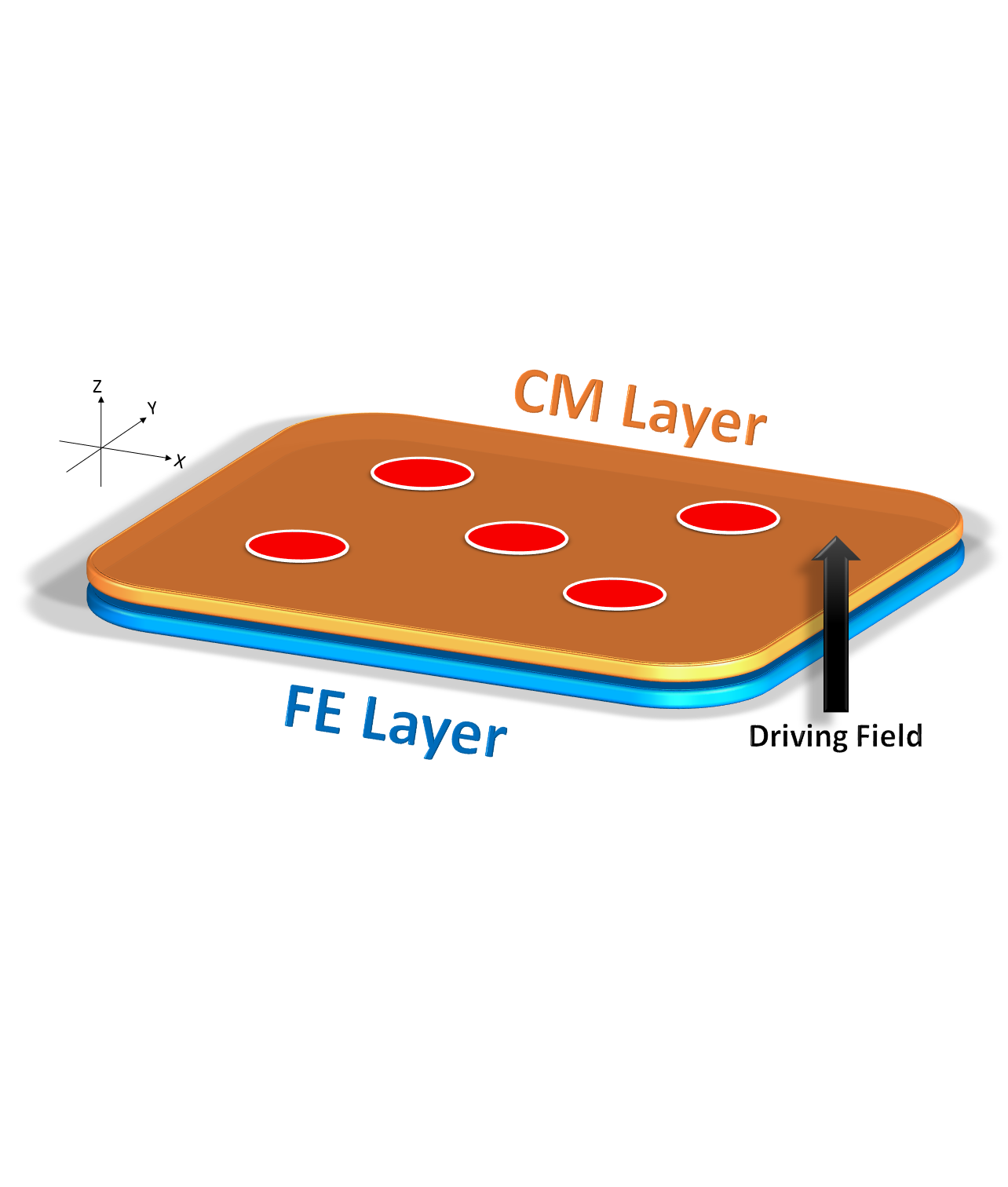}
	\caption{ \textbf{Schematic illustration of a CM/FE bilayer lattice.} The top-plane represents the CM layer, which can form Bloch Skyrmions as shown in the red circle; the bottom-plane represents the FE layer. An external field is applied perpendicularly through the lattice.}
	\label{Fig.1}
\end{figure}

The composite bilayer lattice has been considered as a two-dimensional CM/FE bilayer structure. It consists of $ N \times N $  magnetic spins and electric pseudospins in each layer. The CM layer and the FE layer are glued together by magnetoelectric coupling. Note that each magnetic spin is coupled with an electric pseudospin. The schematic view is shown in \textbf{Figure~\ref{Fig.1}}. Therefore, the total energy $ \mathcal{H} $ for the microscopic model can be written as a sum of three terms: the Hamiltonian in the CM layer $ \mathcal{H}_{CM} $, the Hamiltonian in the FE layer $ \mathcal{H}_{FE} $ and the magnetoelectric interaction between the CM and the FE structure $ \mathcal{H}_{ME} $:

\begin{equation}
\mathcal{H} = \mathcal{H}_{CM} + \mathcal{H}_{FE} + \mathcal{H}_{ME}
\label{Eq.1}
\end{equation}

The spin system in the CM structure can be described by a classical Heisenberg model. The local magnetic spin is represented by $ \textbf{S}_{i,j} = (S_{i,j}^x , S_{i,j}^y , S_{i,j}^z) $, which is normalized, i.e., $ \| \textbf{S}_{i,j} \| = 1 $ , and $ i,j \in [1,2,3,...,N] $  defines the location of the magnetic spin in the lattice. Hence the Hamiltonian $ \mathcal{H}_{CM} $ is given by
\begin{equation}
\mathcal{H}_{CM} = \mathcal{H}_{CM}^{int} + \mathcal{H}_{CM}^{dmi} + \mathcal{H}_{CM}^{ani} + \mathcal{H}_{CM}^{ext}
\label{Eq.2}
\end{equation}
The first term $ \mathcal{H}_{CM}^{int} $ stands for the nearest-neighbor exchange interaction, and $ J_{CM}^* = J_{CM} / k_B T $ is the dimensionless exchange interaction coupling coefficient,
\begin{equation}
\mathcal{H}_{CM}^{int} = -J_{CM} \sum_{i,j} [ \textbf{S}_{i,j} \cdot ( \textbf{S}_{i+1,j} + \textbf{S}_{i,j+1}) ]
\label{Eq.3}
\end{equation}
The second term $ \mathcal{H}_{CM}^{dmi} $ stands for the Dzyaloshinskii-Moriya interaction \cite{JJAP.54.053001}, which is a non-linear exchange interaction that specifies the helicity of the Skyrmions,
\begin{equation}
\mathcal{H}_{CM}^{dmi} = -D_{CM} \sum_{i,j} [ \textbf{S}_{i,j} \times \textbf{S}_{i+1,j} \cdot \hat{x} + \textbf{S}_{i,j} \times \textbf{S}_{i,j+1} \cdot \hat{y} ]
\label{Eq.4}
\end{equation}
where $ D_{CM}^* = D_{CM} / k_B T $ is the dimensionless Dzyaloshinskii-Moriya interaction coefficient, and $ \hat{x} $ and $ \hat{y} $ are the unit vectors of the \textit{x}- and the \textit{y}-axes, respectively. The third term $ \mathcal{H}_{CM}^{ani} $ stands for the magnetic anisotropy,
\begin{equation}
\mathcal{H}_{CM}^{ani} = -K_{CM}^z \sum_{i,j} (S_{i,j}^z)^2
\label{Eq.5}
\end{equation}
where $ K_{CM}^* = K_{CM}^z / k_B T $ is the dimensionless uniaxial anisotropic coefficient in the \textit{z}-direction. The fourth term $ \mathcal{H}_{CM}^{ext} $ stands for the external Zeeman energy,
\begin{equation}
\mathcal{H}_{CM}^{ext} = -\mu_0 \chi_m H_{ext}^z \sum_{i,j} S_{i,j}^z
\label{Eq.6}
\end{equation}
where $ H_{ext}^*(t) = \mu_0 \chi_m H_{ext}^z(t) / k_B T $ is a dimensionless external time-dependent magnetic field, applied perpendicular to the lattice sample along the \textit{z}-direction, $ \mu_0 $ is the magnetic permeability of the classical vacuum, and $ \chi_m $ is the magnetic susceptibility in the CM materials.

We have studied the electric dipoles in the system by the transverse Ising model for the electric pseudospins \cite{JAP.118.124109}. The electric pseudospin in the FE structure is regarded as a vector $ \textbf{P}_{k,l} = (P_{k,l}^x , P_{k,l}^y , P_{k,l}^z) $, and $ k,l \in [1,2,3,...,N] $ characterizes the pseudospin’s location. The Hamiltonian $ \mathcal{H}_{FE} $ of the electric subsystem is given by
\begin{equation}
\mathcal{H}_{FE} = \mathcal{H}_{FE}^{int} + \mathcal{H}_{FE}^{tran} + \mathcal{H}_{FE}^{ext}
\label{Eq.7}
\end{equation}
In the transverse Ising model, only the \textit{z}-component of each electric pseudospin has a contribution to the electric exchange interaction $ \mathcal{H}_{FE}^{int} $, and $ J_{FE}^* = J_{FE} / k_B T $ represents the dimensionless nearest-neighbor interaction coefficient between the electric pseudospins:
\begin{equation}
\mathcal{H}_{FE}^{int} = -J_{FE} \sum_{k,l} [ P_{k,l}^z ( {P}_{k+1,l}^z + {P}_{k,l+1}^z ) ]
\label{Eq.8}
\end{equation}
The second term $ \mathcal{H}_{FE}^{tran} $ stands for the transverse energy, where $ \Omega_{FE}^* = \Omega_{FE}^x / k_B T $  is a dimensionless transverse field in the \textit{x}-direction, which is perpendicular to the Ising \textit{z}-direction \cite{JPC.11.5045}:
\begin{equation}
\mathcal{H}_{FE}^{tran} = -\Omega_{FE}^x \sum_{k,l} P_{k,l}^x
\label{Eq.9}
\end{equation}
The third term $ \mathcal{H}_{FE}^{ext} $ stands for the external energy provided by an applied electric field,
\begin{equation}
\mathcal{H}_{FE}^{ext} = -\epsilon_0 \chi_e E_{ext}^z \sum_{k,l} P_{k,l}^z
\label{Eq.10}
\end{equation}
where $ E_{ext}^*(t) = \epsilon_0 \chi_e E_{ext}^z(t) / k_B T $ is a dimensionless time-related electric field, applied perpendicular to the lattice sample along the \textit{z}-direction, $ \epsilon_0 $ is the electric permittivity of free space, and $ \chi_e $ is the dielectric susceptibility.

The size of an electric pseudospin is different from the classical magnetic spin, since the polarization is defined as the electric dipole moment density in dielectric materials. The dipole moment density $ p $ is proportional to the external electric field $ E_{ext} $ \cite{C.Kittel}:
\begin{equation}
p = \epsilon_0 \chi_e E_{ext}
\label{Eq.11}
\end{equation}
In the pseudospin system, the size of each electric pseudospin is proportional to the magnitude of its effective field $  \| \textbf{E}_{k,l}^{eff} \| $. Hence, 
\begin{equation}
\| \textbf{P}_{k,l} \| = \epsilon_0 \Xi_e \| \textbf{E}_{k,l}^{eff} \|
\label{Eq.12}
\end{equation}
where $ \Xi_e $ is the dimensionless pseudo-scalar susceptibility. As a consequence, electric pseudospins have a variable size as does the behavior of the electric dipoles.

The behavior of the multiferroic is related to the magnetoelectric coupling at the interface, and this can be described by the spin-dipole interaction $ \mathcal{H}_{ME} $ \cite{N.442.759}. The analytic expression of magnetoelectric coupling can be linear or non-linear, particularly with respect to the thermal effect \cite{PRB.92.134424}. In this Article, we only account for low-energy excitations between the CM and FE layers and so we restrict ourselves to the linear expression of the magnetoelectric interaction \cite{PRB.90.054423}, as
\begin{equation}
\mathcal{H}_{ME} = -g \sum_{(i,j)(k,l)} (S_{i,j}^z P_{k,l}^z)
\label{Eq.13}
\end{equation}
where $ g^* = g / k_B T $ is the dimensionless magnetoelectric coupling coefficient. The magnetoelectric coupling was discussed by Spaldin \cite{PRB.93.195167}. The coupling strength $ g $  is, however, unknown. Note that a non-linear expression form has not been studied here, for simplicity, and due to its minor effect in the numerical modeling.

\section{Method}
\label{Method}

The time evolutions of the magnetic spin/electric pseudospin responses are studied by numerically solving the Landau-Lifshitz equations. In the CM lattice, \textbf{Eq. (\ref{Eq.14})} shows a differential equation which predicts the rotation of a magnetic spin in response to its torques (see \textbf{Movie 1} in the Supplemental Material \cite{Supplementary.6}),
\begin{equation}
\frac{\partial \textbf{S}_{i,j}}{\partial t} = -\gamma_{CM} [\textbf{S}_{i,j} \times \textbf{H}_{i,j}^{eff}] - \lambda_{CM} [\textbf{S}_{i,j} \times (\textbf{S}_{i,j} \times \textbf{H}_{i,j}^{eff})]
\label{Eq.14}
\end{equation}
where $ \gamma_{CM} $ is the gyromagnetic ratio which relates the magnetic spin to its angular momentum,  $ \lambda_{CM} $ is the phenomenological damping term in the CM lattice, and $ \textbf{H}_{i,j}^{eff} $ is the effective field of each magnetic spin. This is the derivative of the system Hamiltonian of \textbf{Eq. (\ref{Eq.1})} with respect to the magnitudes of the magnetic spin in each direction, as $ \textbf{H}_{i,j}^{eff} = - \dfrac{\delta \mathcal{H}}{\delta \textbf{S}_{i,j}} $ \cite{JPD.48.305001}.

In the FE layer, the pseudospins describe the locations of the electric dipoles. The electric dipole moment is a measure of the separation of positive and negative charges in the \textit{z}-direction. It is scalar. Consequently, the time evolution of the electric pseudospin is expected to perform a precession free trajectory \cite{JAP.119.124105}, with (see \textbf{Movie 2} in the Supplemental Material \cite{Supplementary.6})
\begin{equation}
\frac{\partial \textbf{P}_{k,l}}{\partial t} = - \lambda_{FE} [\textbf{P}_{k,l} \times (\textbf{P}_{k,l} \times \textbf{E}_{k,l}^{eff})]
\label{Eq.15}
\end{equation}
where $ \lambda_{FE} $ is the phenomenological damping term in the FE structure, and $ \textbf{E}_{k,l}^{eff} $ is the electric effective field for each pseudospin. It is defined as a functional derivative of \textbf{Eq. (\ref{Eq.1})}, as $ \textbf{E}_{k,l}^{eff} = - \dfrac{\delta \mathcal{H}}{\delta \textbf{P}_{k,l}} $.

\section{Skyrmions in an Electric-Square-Field}
\label{E-Skyrmions}

The results have been obtained by solving the Landau-Lifshitz equations in a fourth-order Range-Kutta method with a dimensionless time step for $ \Delta t^* = 0.001 $. In this section, we implement a dimensionless parameter set: $ J_{CM}^* = 1 $, $ D_{CM}^* = 1 $, $ K_{CM}^* = 0.1 $, $ J_{FE}^* = 1 $, $ \Omega_{FE}^* = 0.1 $, $ \Xi_e^* = 0.1 $ , $ g^* = 0.5 $, $ \gamma_{CM}^* = 1 $ and $ \lambda_{CM}^* = \lambda_{FE}^* = 0.1 $. Note that \textquoteleft $ * $\textquoteright s characterize dimensionless quantities in this Article. The numbers of magnetic spins and electric pseudospins are $ N_{CM} = N_{FE} = 23 \times 23 $ arranged in a square lattice. Free boundary conditions and a random initial state are used. The CM/FE bilayer is driven by the electric field only [i.e., no magnetic external field, $ H_{ext}^*(t) = 0 $].

\begin{figure}
	\includegraphics[width=250px, trim=0 40 0 0, clip]{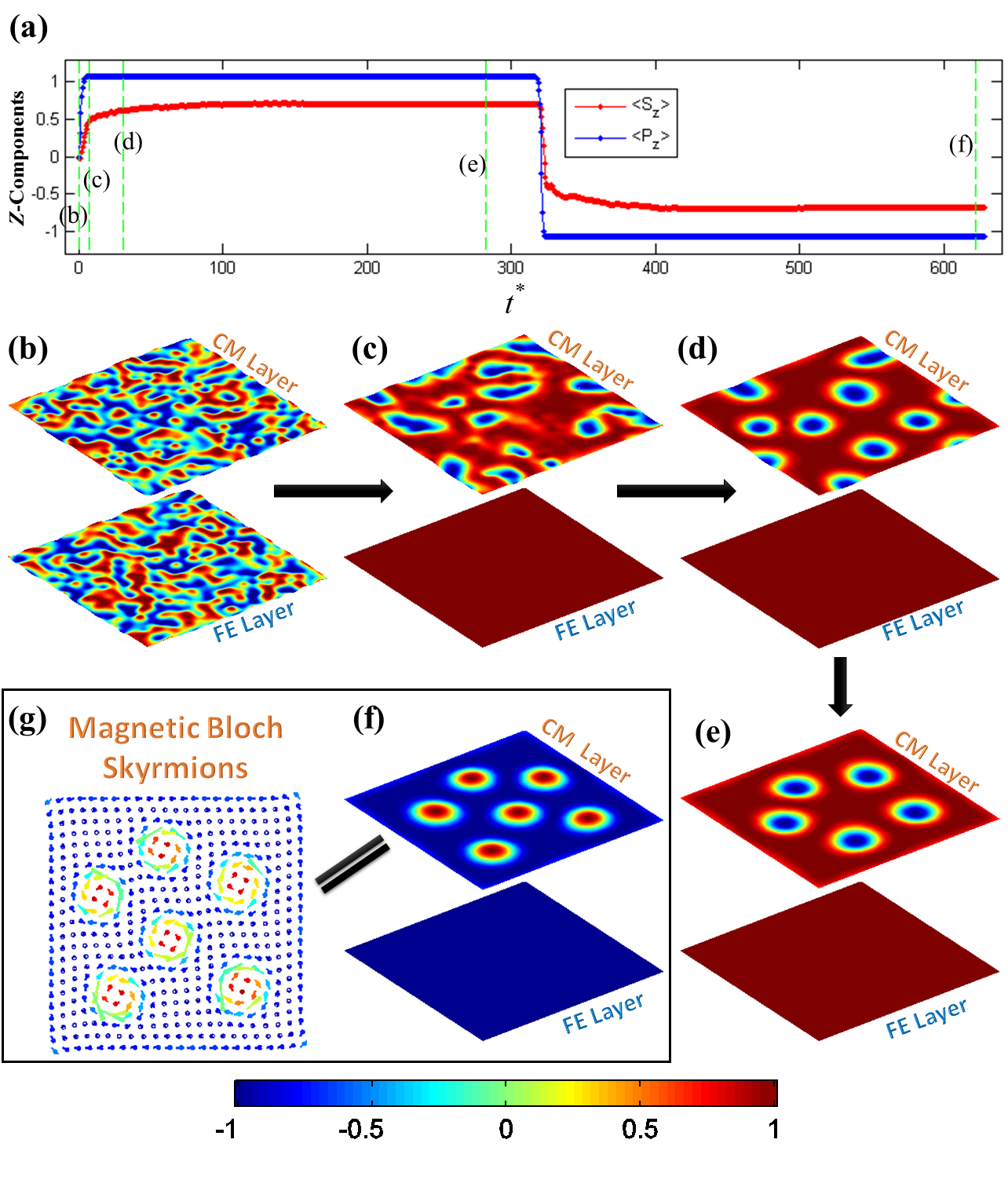}
	\caption{ \textbf{Generation of magnetic Bloch Skyrmions in a CM/FE bilayer by applying an electric square field.}  \textbf{(a)} The mean \textit{z}-components of the magnetization (red) and the electric polarization (blue) to an electric square field. A series of sequential \textbf{(b)$ \rightarrow $(c)$ \rightarrow $(d)$ \rightarrow $(e)} images show the real-time details of magnetic Skyrmion generation. The top-plane is the CM layer, and the bottom-plane is the FE layer. \textbf{(f)} Revised Skyrmions in a different direction of the electric driving field. \textbf{(g)} A top view of magnetic Bloch Skyrmions in the spin-plot. The color scale represents the magnitude of the \textit{z}-component. See \textbf{Movie 3} in the Supplemental Material \cite{Supplementary.6}. }
	\label{Fig.2}
\end{figure}

We obtain the magnetic Bloch Skyrmions induced by an electric field with a square wave form and a dimensionless amplitude $ E_0 = 10 $. \textbf{Figure~\ref{Fig.2}} shows the Skyrmion generation process. The mean out-of-plane components of the magnetization $ S_{z} $ (in the red curve) and the electric polarization $ P_{z} $ (in the blue curve) as functions of time are depicted in \textbf{FIG.~\ref{Fig.2}(a)}, respectively. A series of time evolution images shown the generation progress of the magnetic Skyrmion in \textbf{ FIGs.~\ref{Fig.2}(b)$ \rightarrow $(c)$ \rightarrow $(d)$ \rightarrow $(e) }. An initial state with random configurations on both layers at $ t^* = 0 $ is used [\textbf{FIG.~\ref{Fig.2}(b)}]. Subsequently, the FE structure quickly completes full alignment, but the CM structure orders much more slowly [\textbf{FIG.~\ref{Fig.2}(c)}]. Several baby Skyrmions appear a short time later in \textbf{FIG.~\ref{Fig.2}(d)}. The baby Skyrmions formed around the edge have a very short life. Eventually, five of them in the bulk lattice survive as stabilized Bloch Skyrmions [\textbf{FIG.~\ref{Fig.2}(e)}].

The stabilized Skyrmions require the system to have the lowest free energy. This can be manipulated by the alignment and the size of these Skyrmions. Importantly, the magnetoelectric coupling in the composite bilayer acts as the source of a magnetic driving field in the chiral magnets.  Since each magnetic spin is bound with an electric pseudospin, the magnetic Skyrmions can be manipulated by their related electric polarization. As a consequence, the opposite direction of the electric field [i.e., $ E_{ext}^*(t) = -10 $] reverses the magnetic Skyrmions, as shown in \textbf{FIG.~\ref{Fig.2}(f)}. \textbf{Figure~\ref{Fig.2}(g)} presents a spin-plot of these Bloch Skyrmions in the CM lattice of \textbf{FIG.~\ref{Fig.2}(f)}.

\begin{figure}
	\includegraphics[width=250px, trim=0 40 0 0, clip]{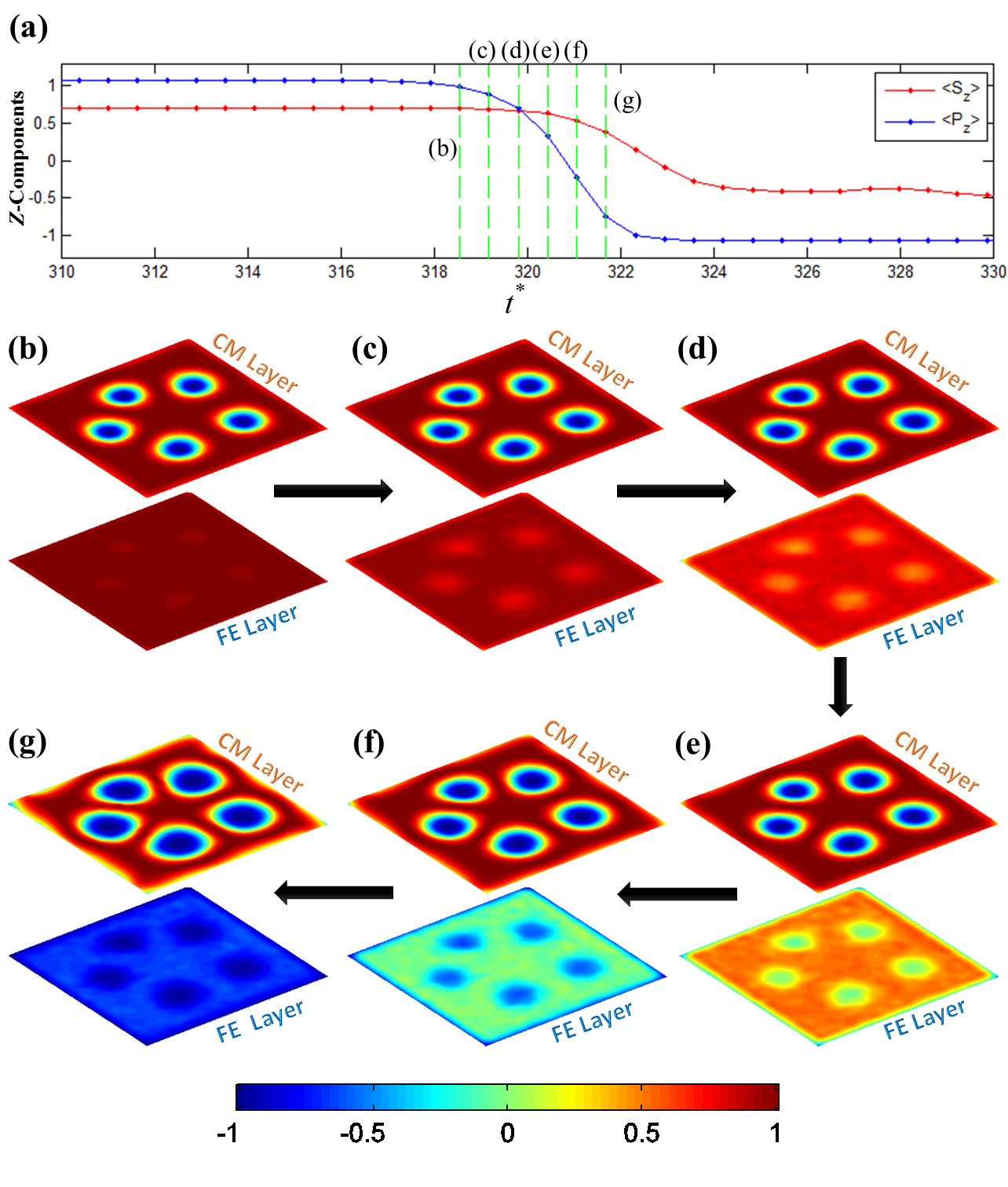}
	\caption{ \textbf{Generation of the electric \textquotedblleft footprint Skyrmions\textquotedblright~during the switching process [i.e., $ E_{ext}^*(t) = 10\rightarrow-10 $].}  \textbf{(a)} The mean \textit{z}-components of the magnetization (red) and the electric polarization (blue). A series of sequential \textbf{(b)$ \rightarrow $(c)$ \rightarrow $(d)$ \rightarrow $(e)$ \rightarrow $(f)$ \rightarrow $(g)} images show the emergence and the vanishing of electric \textquotedblleft footprint Skyrmions\textquotedblright~during the switching process. See \textbf{Movie 4} in the Supplemental Material \cite{Supplementary.6}.}
	\label{Fig.3}
\end{figure}

Interestingly, during the switching process (which introduces as a sudden change of the direction of the electric field, here $ E_{ext}^*(t) = 10\rightarrow-10 $), we observe some skyrmion-like features appearing in the FE layer, as shown in \textbf{FIG.~\ref{Fig.3}}. \textbf{Figure~\ref{Fig.3}(a)} presents the mean out-of-plane components of the magnetization $ S_z $ (in the red curve) and the electric polarization $ P_z $ (in the blue curve) in a limited time zone, $ t \in (310 , 330) $, which is taken around the switching of the electric square field. A sequence of images in \textbf{ FIGs.~\ref{Fig.3}(b)$ \rightarrow $(c)$ \rightarrow $(d)$ \rightarrow $(e)$ \rightarrow $(f)$ \rightarrow $(g) } shows the magnetic Skyrmions are consistent with the Skyrmion footprints in the FE structure. Hence, we call them the \textit{electric \textquotedblleft footprint Skyrmions\textquotedblright}. They only appear for an extremely short period during the reorienting process.

To understand why the electric \textquotedblleft footprint Skyrmions\textquotedblright~can be created, we consider the \textit{footprint effect}. During the switching process, the electric exchange interaction energy $ \mathcal{H}_{FE}^{int} $ is used to support electric pseudospins responding to the driving field. Therefore, as the reduction of $ \mathcal{H}_{FE}^{int} $, the binding energy (magnetoelectric interaction $ \mathcal{H}_{ME} $) dominates the system.  The magnetic Skyrmions provide a non-uniform magnetization distribution in the CM structure. This indicates that different binding energies in each spin-pseudospin bond result in the \textquotedblleft footprints\textquotedblright~of the FE structure. These \textquotedblleft footprints\textquotedblright~exist for a limited time, since the electric pseudospins reorient quickly in response to the driving field. Eventually, since $ \mathcal{H}_{FE}^{int} $ has been saturated, the \textquotedblleft footprints\textquotedblright~disappear, as shown in \textbf{FIG.~\ref{Fig.3}(g)}.  Consequently, we can control the electric \textquotedblleft footprint Skyrmions\textquotedblright~by the driving-field application. To confirm this, in the next section we demonstrate that stabilized electric \textquotedblleft footprint Skyrmions\textquotedblright~can be induced by a magnetic field.

\section{Electric \textquotedblleft Footprint Skyrmions\textquotedblright~in a Magnetic Field}
\label{M-Skyrmions}

One of our aims in this Article is to obtain strong electric \textquotedblleft footprint Skyrmions\textquotedblright. So far, we have observed temporary \textquotedblleft footprint Skyrmions\textquotedblright~in the FE layer, which occur during the switching process of the electric field. Now, we replace the electric square field by a static magnetic driving field, in order to enhance the stability of the electric \textquotedblleft footprint Skyrmions\textquotedblright. A dimensionless parameter set is selected:  $ J_{CM}^* = 1 $, $ D_{CM}^* = 1 $, $ K_{CM}^* = 0.1 $, $ J_{FE}^* = 0.01 $, $ \Omega_{FE}^* = 0.1 $, $ \Xi_e^* = 1 $ , $ g^* = 0.5 $, $ \gamma_{CM}^* = 1 $ and $ \lambda_{CM}^* = \lambda_{FE}^* = 0.1 $. To study the magnetic-field-driven dynamics, the same lattice sample $ N_{CM} = N_{FE} = 23 \times 23 $, free boundary conditions, and a static magnetic field  $ H_{ext}^*(t) = 0.5 $ are used.

\begin{figure}
	\includegraphics[width=250px, trim= 0 450 0 0, clip]{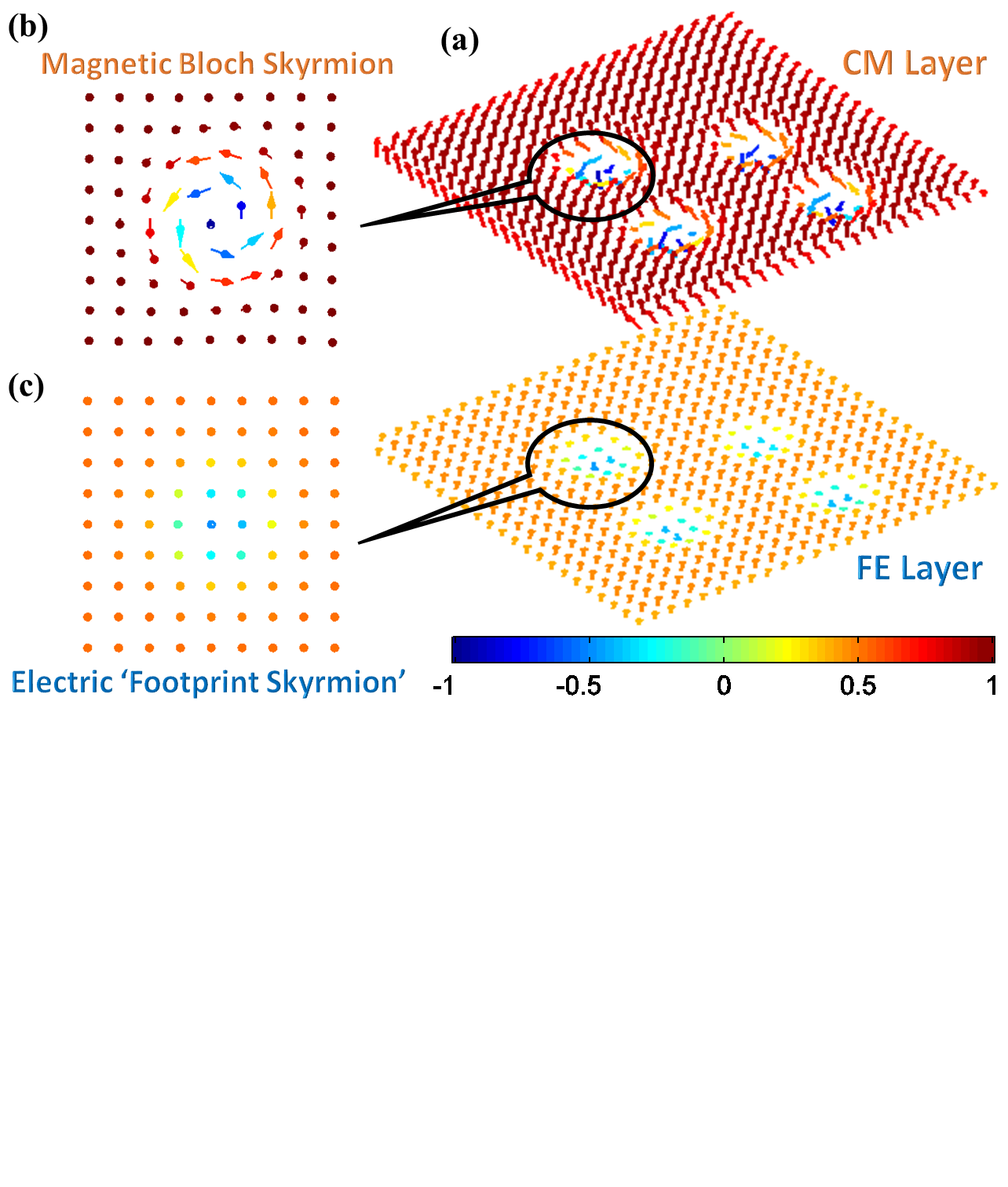}
	\caption{ \textbf{(a)} Magnetic Skyrmions and induced electric \textquotedblleft Skyrmions\textquotedblright~to a magnetic driving field. Closer top-views of a magnetic Bloch Skyrmion in \textbf{(b)} and an electric \textquotedblleft footprint Skyrmion\textquotedblright~in \textbf{(c)}. The color scales represent the magnitudes of the \textit{z}-component. See \textbf{Movie 5} in the Supplemental Material \cite{Supplementary.6}.}
	\label{Fig.4}
\end{figure}

\textbf{Figure~\ref{Fig.4}} demonstrates that the electric \textquotedblleft footprint Skyrmions\textquotedblright~can be induced by a magnetic driving field in the stacked CM/dielectric bilayer lattice. Magnetic Bloch Skyrmions are formed in the CM structure due to their direct response to the magnetic driving field. Since the electric pseudospins are coupled with their related magnetic spins by magnetoelectric coupling, hence the \textquotedblleft footprint Skyrmions\textquotedblright~are projected in the dielectric lattice. A closer inspection shows a magnetic Bloch Skyrmion with the spiral texture in \textbf{FIG.~\ref{Fig.4}(b)}, but the electric \textquotedblleft footprint Skyrmion\textquotedblright~does not have one, as shown in \textbf{FIG.~\ref{Fig.4}(c)}. That is due to pseudospins representing the electric dipoles. The electric dipoles only have magnitudes, but no directions. Therefore, the electric pseudospins located in the \textquotedblleft footprint Skyrmion\textquotedblright~have different magnitudes to pseudospins elsewhere. Hence, the \textquotedblleft footprint Skyrmions\textquotedblright~can only be observed by the different colors. In the absence of the electric field and with a weak electric exchange coupling ($ J_{FE}^* = 0.01 $), the sizes of pseudospins are finite [see \textbf{Eq. (\ref{Eq.12})}]. Thus, the color in the dielectric lattice is lighter than the CM lattice. This type of electric \textquotedblleft footprint Skyrmions\textquotedblright~is consistent with the corresponding magnetic Skyrmions.

\begin{figure}
	\includegraphics[width=250px, trim= 0 400 0 0, clip]{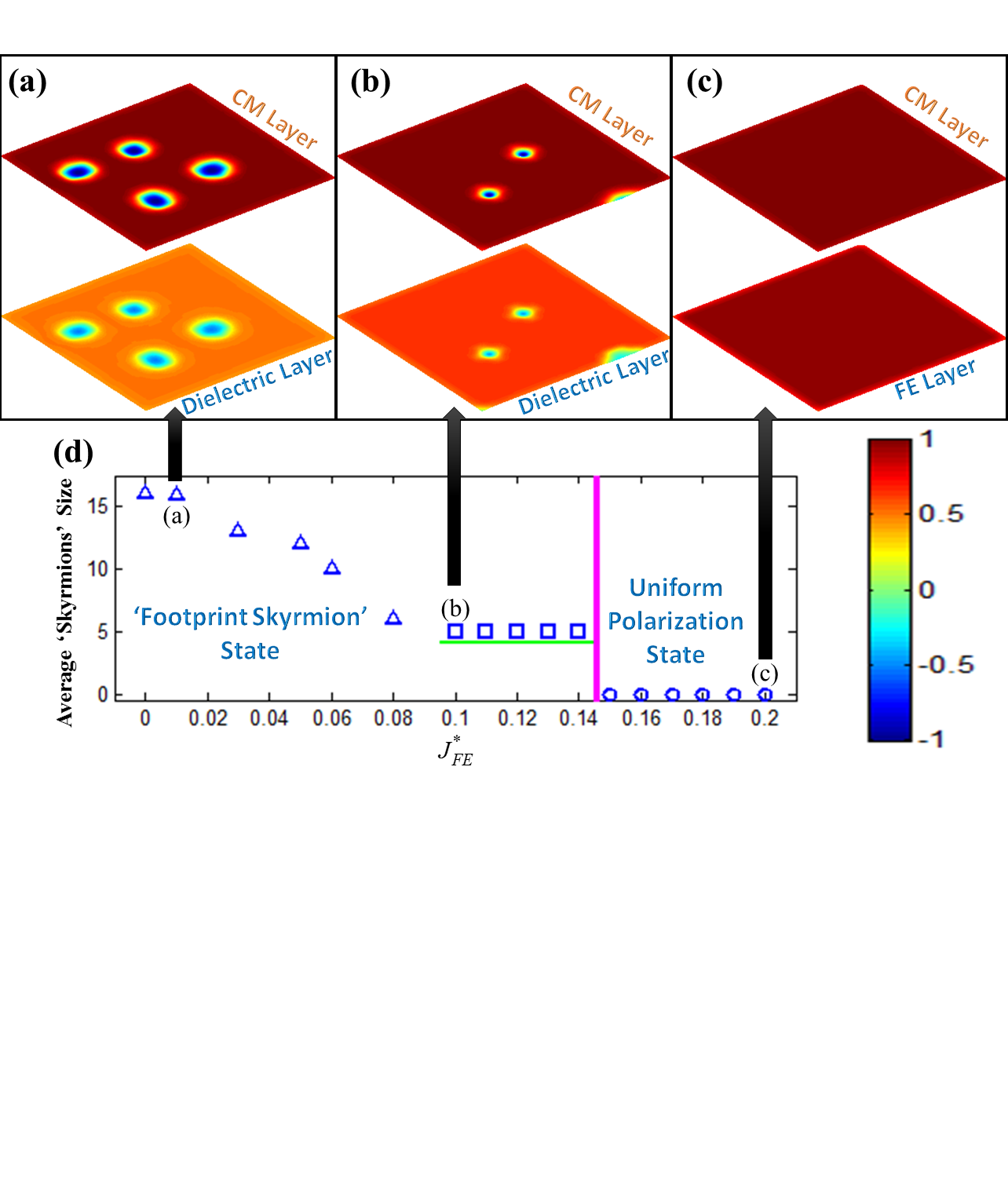}
	\caption{ \textbf{A comparison of the electric exchange coupling effects to the Skyrmions in the bilayer.}  \textbf{(a)} $ J_{FE}^* = 0.01 $, \textbf{(b)} $ J_{FE}^* = 0.1 $, and \textbf{(c)} $ J_{FE}^* = 0.2 $. The color scale represents the magnitude of the \textit{z}-component. \textbf{(d)} Average sizes of various footprints with the electric exchange couplings $ J_{FE} $.}
	\label{Fig.5}
\end{figure}

In \textbf{Section~\ref{E-Skyrmions}}, we have discussed that electric \textquotedblleft footprint Skyrmions\textquotedblright~can be created during the switching process, when the magnetoelectric interaction dominates the energy in the FE structure. It is the key feature to obtain electric \textquotedblleft footprint Skyrmions\textquotedblright. In this section, the electric external field is absent, and the transverse energy is small. Thus, the electric exchange interaction only competes with the magnetoelectric interaction, since the electric exchange interaction is modulated by the electric exchange coupling $ J_{FE} $  in the simulation. We compare the sizes of Skyrmions with the different magnitudes of  $ J_{FE} $ under a constant magnetic field in \textbf{FIG.~\ref{Fig.5}}. \textbf{Figure~\ref{Fig.5}(a)} with $ J_{FE}^* = 0.01 $ contains four electric \textquotedblleft Skyrmions\textquotedblright~with their corresponding magnetic Skyrmions. As $ J_{FE} $ increases, both magnetic Skyrmions and electric \textquotedblleft Skyrmions\textquotedblright~reduce their size. As seen in \textbf{FIG.~\ref{Fig.5}(b)} with $ J_{FE}^* = 0.1 $, two tiny Skyrmions and two \textquotedblleft footprint Skyrmions\textquotedblright~have survived. Finally, neither Skyrmions nor \textquotedblleft footprint Skyrmions\textquotedblright~survive, as observed in \textbf{FIG.~\ref{Fig.5}(c)} with $ J_{FE}^* = 0.2 $. This can be traced back to the coupling between the CM layer and the FE layer by the magnetoelectric effects. \textbf{Figure~\ref{Fig.5}(d)} summarizes the effect of the \textquotedblleft footprint Skyrmions\textquotedblright~size to the strength of the electric exchange couplings. Here, the size is measured by counting the number of pseudospins contributing to the \textquotedblleft footprint Skyrmions\textquotedblright. Results of the numerical simulations show the weaker electric exchange couplings (i.e., $ J_{FE}^* \in [0 , 0.14] $) can conserve the \textquotedblleft footprint Skyrmions\textquotedblright. Otherwise, the uniform polarization appears due to the electric exchange interaction energy dominating the energy contribution in the FE system, as shown in \textbf{FIG.~\ref{Fig.5}(c)}. Minimal \textquotedblleft footprint Skyrmions\textquotedblright~contain five pseudospins (i.e., one center and four neighbors), which have been detected as $ J_{FE}^* \in [0.1 , 0.14] $. In this case, the polarization in these \textquotedblleft Skyrmions\textquotedblright~decreases as $ J_{FE} $ increases. More details are shown in the Supplemental Material \textbf{Figure 1} \cite{Supplementary.6}.

\section{Conclusion}
\label{Conclusion}

We have demonstrated that magnetic Bloch Skyrmions in the CM structure can be induced by an electric field. Interestingly, since the FE layer is coupled with the chiral magnet by the magnetoelectric effect, it offers an opportunity for magnetic Skyrmions in the CM layer to produce projections onto the FE structure. We call these projections \textquotedblleft footprint Skyrmions\textquotedblright. Electric \textquotedblleft footprint Skyrmions\textquotedblright~can be generated by either an electric field, or a magnetic field. In the electric driving field, the \textquotedblleft footprint Skyrmions\textquotedblright~only exist during the switching of the electric field. In the magnetic driving field, the \textquotedblleft footprint Skyrmions\textquotedblright~are stable.

\begin{acknowledgments}
	The authors thank L. Chotorlishvili for discussions. Wang Zidong gratefully acknowledges Wang Yuhua, Zhao Bingjin, Zhao Wenxia, and Wang Feng for support.
\end{acknowledgments}

\bibliography{Bibliography}

\end{document}